Temperature dependence of electron-spin relaxation in a single InAs quantum dot at zero applied magnetic field

X. M. Dou, B. Q. Sun[*], D. S. Jiang, H. Q. Ni, and Z. C. Niu
SKLSM, Institute of Semiconductors, CAS, P. O. Box 912, Beijing 100083, China

The temperature-dependent electron spin relaxation of positively charged excitons in a single InAs quantum dot (QD) was measured by time-resolved photoluminescence spectroscopy at zero applied magnetic fields. The experimental results show that the electron-spin relaxation is clearly divided into two different temperature regimes: (i) $T < 50$ K, spin relaxation depends on the dynamical nuclear spin polarization (DNSP) and is approximately temperature-independent, as predicted by Merkulov et al. (ii) $T >$ about 50 K, spin relaxation speeds up with increasing temperature. A model of two LO phonon scattering process coupled with hyperfine interaction is proposed to account for the accelerated electron spin relaxation at higher temperatures.

[*] Electronic mail: bqsun@semi.ac.cn

## I. INTRODUCTION

Semiconductor quantum dots (QDs), a natural candidate for the qubit operations of quantum computing, have attracted great attention due to that the localized character of the electron wave functions in the QDs suppresses the most effective intrinsic spin-flip mechanisms related to the absence of inversion symmetry in GaAs-like crystals.[1] This leads to an unusual low rate of spin-flip transitions. Recently, however, theoretical and experimental studies have shown that the dominant mechanism of electron-spin relaxation in the QDs at low temperature is due to the hyperfine interaction with randomly oriented nuclear spins,[2-5] which gives rise to a decay time of the order of *ns* for an ensemble of QDs. However, this process will be strongly suppressed by applying external magnetic field or using photo-oriented electrons in the QDs to polarize nuclear spins as what occurs during the experiment of polarized photoluminescence (PL).[4-7] In order to measure electron-spin relaxation under different conditions of the dynamical nuclear spin polarization (DNSP) it is important to tune DNSP in the experiment and to check the influence of DNSP on the spin-flip process.[5,8-11] On the other hand, when the temperature is raised up, the possible phonon-assisted spin-flip transitions via hyperfine interaction,[1,12-14] spin-orbit coupling[1,12,15,16], or other mechanisms [1,17] have been widely investigated in theory, but only several experimental works were reported.[18-21] Therefore, it is important to check the mechanisms of the temperature-dependent spin relaxation experimentally by tuning DNSP furthermore.

In this article, we have investigated the temperature-dependent electron-spin relaxation (characterized by $T_1$ time) in a single InAs QD with positively charged exciton ($X^+$) by time-resolved photoluminescence (TRPL) spectroscopy at zero applied magnetic fields. The randomly oriented or polarized nuclear spins were created by alternating $\sigma^+/\sigma^-$ or $\sigma^+$ pulsed sequences, respectively, to excite the QDs. The experimental results show that electron-spin relaxation is divided into two different temperature regimes: (i) $T < 50$ K, spin relaxations are approximately temperature-independent and dominated mainly by the DNSP as predicted by Merkulov et al.[3] (ii) $T >$ about 50 K, spin relaxation speeds up with increasing temperature. A model of two LO phonon scattering process coupled with hyperfine interaction at higher temperatures is proposed to qualitatively account for electron spin relaxation. In addition, the temperature-dependence of the Zeeman splitting due to internal magnetic field instead of external magnetic field was checked under $\sigma^+$ excitation.

## II. EXPERIMENTAL

The investigated QD samples were grown by molecular beam epitaxy on a semi-insulating GaAs substrate. They consist of, in sequence, an *n*-doped GaAs buffer layer, a 20-period *n*-doped GaAs/Al$_{0.9}$Ga$_{0.1}$As distributed Bragg reflector (DBR), a 2λ GaAs cavity with an InAs QD layer at the cavity antinode, and a top *p*-doped GaAs layer. An ultra-low density of InAs QD layer was formed by depositing nominally 2.35 monolayers (ML) of InAs at a growth rate of 0.001 ML/s. In experiments, the QD sample was mounted in a continuous-flow liquid helium cryostat and measured under temperatures from 10 to 70 K. A mode-locked Ti: sapphire laser with 2 ps pulses and 80 MHz repetition frequency was tuned to a wavelength of 902 nm to excite the QD sample by the GaAs LO-phonon-assisted resonance excitation.[5] The excitation intensity is about 5 μW. The emission line of $X^+$ from the positively charged exciton has been identified and reported previously in Ref. 22. The emitted luminescence was collected by an objective (NA: 0.5), spectrally filtered by a 0.5 m monochromator, and then detected by a silicon charge coupled device (CCD). For measuring high-resolution PL spectra (HRPL), a scanning Fabry-Perot interferometer (FPI) with free spectral range of 15 GHz (62 μeV) and spectral resolution of 3μeV, a multi-channel scaler (MCS), and an avalanche photodiode (APD) were used. TRPL measurements were carried out by a time-correlated single-photon counting (TCSPC) setup with a time resolution of 400 ps. The time-dependence of both $\sigma^+$ and $\sigma^-$ components of the polarization PL is measured after the $\sigma^+$ excitation pulse. In order to control the nuclear spins in the QD, two different kinds of excitation pulse sequences are used. In the first configuration, the pulses are alternating with $\sigma^+$ and $\sigma^-$ (T-arm and R-arm, respectively, as shown in Fig.1) polarizations, and these pulses are separated by 6.25 ns to excite the QD sample. In this configuration, the nuclear spins are not polarized as two opposite pulses are working alternately. However, in another configuration, the nuclear spins are polarized when only one pulse sequence with $\sigma^+$ polarization (T-arm in Fig. 1) excites the QD sample. This setup is first employed for the TRPL measurements of a single QD,[5] and the mechanism of temperature-dependent electron-spin relaxation under the influence of either random or polarized orientation of nuclear spins is tested in this article.

III. RESULTS AND DISCUSSION

For the positively charged exciton $X^+$ in a single QD, two holes form a spin singlet and the unpaired single electron interacts with the nuclei during the radiative lifetime of the excitonic recombination in about 1 ns. The PL measurements of the circular polarization degree ($P_c$) of the $X^+$ emission in the QDs following circularly polarized laser excitation in such a time scale thus will directly probe the spin polarization of the electron as $\langle S_z^e \rangle = -P_c/2$. Time-dependent circular polarization degree can be derived by the expression $P_c = (I_{\sigma+} - I_{\sigma-})/(I_{\sigma+} + I_{\sigma-})$, where $I_{\sigma+}$ and $I_{\sigma-}$ are the emission intensities of TRPL with $\sigma^+$ and $\sigma^-$ components. At low temperature the electron-spin relaxation processes significantly depend on the DNSP, exhibiting a fast electron-spin relaxation with either a decay time of 0.55 ns at randomly oriented nuclear spins, or a decay time prolonging to 34 ns under polarized nuclear spins. These results have been discussed in details in Ref. 5.

The time-dependent $P_c$ was measured at different temperatures between 10 and 70 K under either only $\sigma^+$ or alternating $\sigma^+/\sigma^-$ pulse sequence excitation. Figures 2 (a) and (b) depict the $P_c(t)$ decay curves obtained at 10, 40, 60, and 70 K under the two excitation configurations, respectively. It clearly shows in a semi-logarithmic scale of the time-dependent $P_c$ that a decay process is much faster for $T = 60$ and 70 K, and the curves can be well described by a single exponential decay, whereas for $T = 10$ and 40 K, especially for the case of $\sigma^+/\sigma^-$ pulse sequence excitation shown in Fig.2 (b), the decay curves are rather nonlinear ones which has already been discussed both in theory and experimentally.[3,5] However, the decay curves at such as 10 K and 40 K in Fig.2 (b) can be approximately divided into two linear parts, i.e. a slower one in the time range of 0-0.5 ns and a faster one in the time range of 0.5-1.5 ns. We take the latter time range to exponentially fit the decay curves and get decay time $T_1$ for $T \leq 40$ K. However, to get a more precise decay time value, the deconvolution calculations of the decay curve should be employed as the decay time is close to the time-resolution of TCSPC. This is valid for all experimental data of Fig. 2 (b) in the case of alternating $\sigma^+/\sigma^-$ pulse excitation, and also for the data of Fig. 2 (a) in the temperature range of 55-70 K under $\sigma^+$ pulse excitation. The obtained temperature dependences of spin relaxation rate $1/T_1$ from 10 to 70 K are plotted in Fig. 3, where open and solid squares correspond to the results with either $\sigma^+/\sigma^-$ or $\sigma^+$ pulse sequence excitation, respectively. It can be seen that when the temperature is lower than 50 K, the spin relaxation rate $1/T_1$ is in the range of 1.3 – 2.2 ns$^{-1}$ for randomly oriented nuclear spins with $\sigma^+/\sigma^-$ excitation, and is reduced as small as about 0.08 ns$^{-1}$ for oriented nuclear spins with $\sigma^+$ excitation, indicating that the spin relaxation has been suppressed in the latter case. However, at elevated temperatures the spin relaxation rate $1/T_1$ increases fast with increasing temperature. In fact, when the temperature is higher than about 50 K, in both excitation configurations the spin relaxation rate increases in a similar way as shown in Fig.3. It suggests that at $T > 50$ K the spin relaxation is no more dominated only by the randomly distributed frozen fluctuation of the nuclear field, instead, it is more likely to become related to a temperature-dependent phonon-assisted process. In this case, the spin relaxation rate $1/T_1$ can be written as the sum of two items: $1/T_1 = 1/T_\Delta + 1/T_{ph}$, where $T_\Delta$ and $T_{ph}$ are decay times due to the hyperfine interaction and phonon-assisted process, respectively. Merkulov et al. have shown that decay time $T_\Delta$ can be written as,[3]

$$T_\Delta = \hbar[n^2 \sum I^j(I^j+1)(A^j)^2/(3N)]^{-1/2} \quad (1)$$

where $N$ is the number of nuclei interacting with the electron in QD, $A^j$ is the hyperfine constant, $I^j$ is the spin of the $j$th nucleus, and $n$ is the number of nuclei per unit cell. The sum goes over all the atoms in the primitive unit cell. Based on Eq. (1), the decay time $T_\Delta$ can be estimated by using the parameters of the hyperfine constants of As ($I^{As} = 3/2$) and In ($I^{In} = 9/2$) nuclei, $A_{As} = 47$ $\mu$eV and $A_{In} = 56$ $\mu$eV, and $n = 2$. For $N \sim 10^5$, Eq. (1) yields $T_\Delta \sim 0.58$ ns for an InAs QD as indicated by the red horizontal line in Fig.3. The well agreement between experimental data and calculated value confirms further that the decay time at $T < 50$ K is determined by the randomly oriented frozen nuclear spins. In this case, it means that at zero external magnetic fields the electron spin relaxation, described by $T_1$ time, is really originated from the dephasing process under frozen fluctuation of the nuclear field, which is not related with any population transfer and is denoted as $T_\Delta$ by Merkulov et al.[3] In addition, it is noted that there is also a very small decrease of $1/T_1$ when $T$ increases from 25 to 40 K, which is probably related to the variation of the temperature-dependent correlation time between electron and nuclei.[23]

On the other hand, for $T >$ about 50 K, the spin relaxation rate speeds up with increasing $T$. Note that as phonon scattering probability increases with increasing $T$, it is reasonable to attribute the observed temperature dependence of $1/T_1$ to a phonon-assisted process. Actually, several important phonon-assisted processes have been proposed to explain the electron spin relaxation mechanisms in the QDs in recent years, such as electron-phonon scattering combined with the spin-orbit (SO) interaction,[1,12,15,16,24-26] two LO-phonon scattering process within the radiative doublet via the first excited state,[17] and phonon-assisted hyperfine interaction between Zeeman sublevels of single electron ground state.[12-14] Some of them could be excluded according to the analysis of the experimental results of InAs QDs. For example, the electron-phonon scattering combined with spin-orbit (SO) interaction normally is only a weak effect on electron-spin relaxation in the QD,[18,24-26] and thus could be excluded. It is noted that the temperature dependence of spin relaxation rate $1/T_1$ fitted to the data in Fig. 3 can be expressed as exp $(-\Delta E/k_B T)$. It gives rise to an activation energy of $\Delta E = (20\pm1)$ meV. This value is consistent with the LO phonon energy of InAs QD of 30 meV taking account of that the bandwidth of the LO phonon is about 8 meV due to disorder fluctuations of the phonon energy.[27] In addition, the second-order electron-LO phonon scattering process in which one phonon is absorbed and another one is emitted becomes more efficient at elevated temperatures.[1,17] In this case, a nonlinear temperature dependence of the relaxation rate related with two-phonon process can be expected, as shown by the temperature dependence of $1/T_1$ in Fig.3 in the temperature range where $k_B T$ is smaller than the LO phonon energy $\hbar\Omega$.[17] The spin relaxation induced by two LO phonon-assisted spin-flip at higher $T$ should be in proportion to the multiplication of number of phonons,[17] i.e., $1/T_1(T) \propto N_{LO}(N_{LO}+1)$, where $N_{LO} = (e^{\hbar\Omega/k_B T} - 1)^{-1}$, $\hbar\Omega$ is LO phonon energy of InAs QD. Actually, the temperature dependence of spin relaxation rate induced by two LO phonon scattering process can be written as,

$$1/T_1(T) = 1/T_1(0) + \alpha N_{LO}(N_{LO}+1) \quad (2)$$

where $T_1(0)$ is temperature-independent spin decay time, and $\alpha$ is the sum of electron-phonon transition rate. By fitting Eq.(2) to the experimental data (solid squares) in Fig. 3, it yields an LO phonon energy $\hbar\Omega = (19.7\pm1.4)$ meV and spin relaxation rate $1/T_1(0) = (0.07\pm0.03)$ ns$^{-1}$. The derived $\hbar\Omega$ can be taken to be well consistent with the LO phonon energy of InAs QD due to that there is a certain bandwidth of the LO phonon and the electron level is broadened at elevated temperatures. In addition, it is found that the derived $1/T_1(0)$ in Eq. (2) is very close to the spin relaxation rate of 0.08 ns$^{-1}$ at 10 K, suggesting that Eq.(2) can semi-quantitatively describe the temperature-dependent electron spin relaxation process in QDs.

Note that electron phonon interaction does not couple with pure electron spin states, but the coupling can be realized through the spin-orbital or nuclear hyperfine interaction. The spin-orbital interaction has only a weak effect on electron-spin relaxation in the QD.[18,24-26] Thus, the hyperfine interaction is probably responsible for the speeding up of spin relaxation at increasing temperature when $T \gtrsim 50$ K, where the hyperfine coupling with nuclei is modulated by lattice vibrations [1,28] via two LO phonon scattering process, or it is a trion-relevant two-phonon-like scattering via fluctuating nuclear fields.[12] However, the more detailed theoretical calculation is needed to thoroughly clarify the temperature dependent spin relaxation under influence of nuclear fields.

Next, in order to obtain the temperature-dependence of the Zeeman splitting ($\Delta E$) at zero external magnetic field, HRPL spectra of the $\sigma^+$ and $\sigma^-$ components under $\sigma^+$ excitation are measured between 10-50 K as shown in Fig.4 (a) for $T$ = 10, 35 and 45 K. Unfortunately, it is found that the full width at half maximum (FWHM) of PL becomes too broadened to clearly separate two peaks at increasing $T$. Hence, it is difficult to measure the Zeeman splitting further after $T$ > 50 K due to the broadening of PL peak and the decrease of PL intensity. But it is still possible to clearly recognize a decrease of $\Delta E$ at higher $T$. The detailed temperature-dependence of Zeeman splitting and FWHM are shown in Fig.4 (b), where the value of FWHM at 50 K is found to be much larger than $\Delta E$. Therefore, at $T \geqslant$ 50 K the Zeeman splitting does exist, but it becomes much smaller as compared to the level broadening at higher $T$. Such a condition will increase the transition probability of second-order electron-LO phonon scattering. At the same time, the fluctuating nuclear field induced by lattice vibrations will also occur at elevated temperatures, leading to a further enhancement of spin relaxation.

Now we compare our work with an interesting experimental work where the temperature-dependent decoherence and dephasing times of ensemble QDs, which correspond to transverse time $T_2$ and $T_2^*$, are investigated, respectively.[18] In Ref. 18 the reported time $T_2$ (~500 ns) at low temperature is close to the predicted time scale given by $\hbar N/A$ (on the order of $\mu s$) for single InAs quantum dot.[2,29] It is assumed that the effects of inhomogeneous broadening are removed by means such as mode locking technique or spin echoes experimentally.[18,30,31] The dephasing time of $T_2^*$ reported follows the temperature dependence similar to that of the spin relaxation of $T_1$ measured in our case because these spin relaxation times are relevant to each other.[18,29,32,33]. Actually, Merkulov et al[3] has predicted that at low temperature and zero magnetic field $T_2^* \approx T_1$ in the QDs, which is experimentally confirmed by Braun et al[4] and Dou et al[5], reflecting that electron dephasing and spin-flip processes are all originated from the same source of the nuclear-spin fluctuations, i.e., they come from the precession of the electron spin in the hyperfine field of the frozen fluctuation of the nuclear spins. In Ref. 18, electron makes a precession at a fixed applied transverse magnetic field，and the transverse time $T_2$ is then interpreted by the theory of modulations of hyperfine field via the phonon-assisted transitions between excited and ground states.[28] In our experiment, however, electron makes a precession at the nuclear field of a snapshot of the frozen fluctuations. The time $T_1$ measured at low temperature thus can be described by the Merkulov process, and at higher temperatures $T_1$ is determined by two LO phonon scattering process via the modulation of the hyperfine interaction.

## IV. SUMMARY

In conclusion, by using alternating $\sigma^+/\sigma^-$ or $\sigma^+$ pulse sequences to optically excite the QD, either randomly oriented or polarized DNSP is generated. This enables us to investigate the temperature-dependent electron-spin relaxation under different configurations of DNSP. The temperature dependence of spin relaxation shows that there are two different $T$ regimes: (i) $T$ < 50 K, spin relaxations are approximately temperature-independent and dominated mainly by the DNSP, which were predicted by Merkulov et al.[3] (ii) $T$ > about 50 K, spin relaxation speeds up. The obtained activation energy of the temperature dependence is consistent with the LO phonon energy, suggesting that the second-order LO phonon-assisted scattering process is related with the electron spin relaxation.

ACKNOWLEDGMENTS

This work is supported by the National Basic Research Program of China and NSPC under Grant Nos. 2007CB924904 and 90921015.

Figure Captions

Fig. 1. (colored online). Schematic diagram of experimental setup of TRPL in which the T-arm and R-arm represent the optical path of alternating $\sigma^+$ and $\sigma^-$ excitation pulse sequences, separated by 6.25 ns. The 2 ps pulses are generated by 100 fs laser pulses going through a monochromator. TAC: time-amplitude converter. MCA: multichannel analyzer. MCS: multichannel scaler.

Fig. 2. (colored online) Time-dependent circular polarization $P_c$ obtained at different temperatures. The results shown here for $T$ = 10 (black), 40 (red), 60 (green), and 70 K (blue), and plotted in a semilogarithmic scale, correspond to either $\sigma^+$ (a) or $\sigma^+/\sigma^-$ (b) pulse sequence excitation.

Fig. 3. (colored online) Temperature-dependences of spin relaxation rate $1/T_1$ under $\sigma^+/\sigma^-$ (open squares) and $\sigma^+$ (solid squares) pulse sequence excitation. The red line is a calculated result from Eq. (1), corresponding to the spin relaxation due to the interaction with the nuclear hyperfine interaction (HF). Blue line is a result of Eq.(2) fit to the experimental data, showing temperature-dependent spin relaxation by two LO phonon-assisted scattering process.

Fig. 4. (colored online) (a) Temperature-dependence of the $\sigma^+$ (red) and $\sigma^-$ (black) components of HRPL spectra under $\sigma^+$ excitation measured by FPI and MCS, where the results of 10, 35, and 45 K plotted in a semilogarithmic scale are shown. The free spectral range of FPI is 62 $\mu$eV. (b) Temperature-dependences of the Zeeman splitting (open squares) and FWHM (open circles) derived from the HRPL spectra.

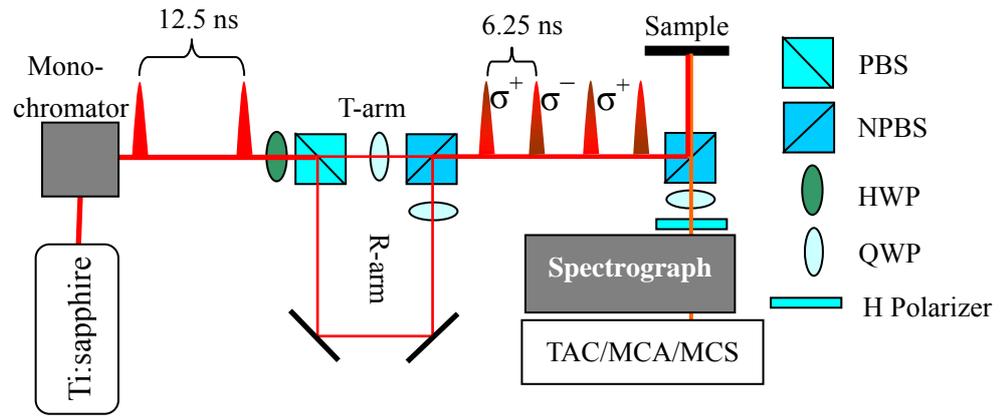

Figure 1

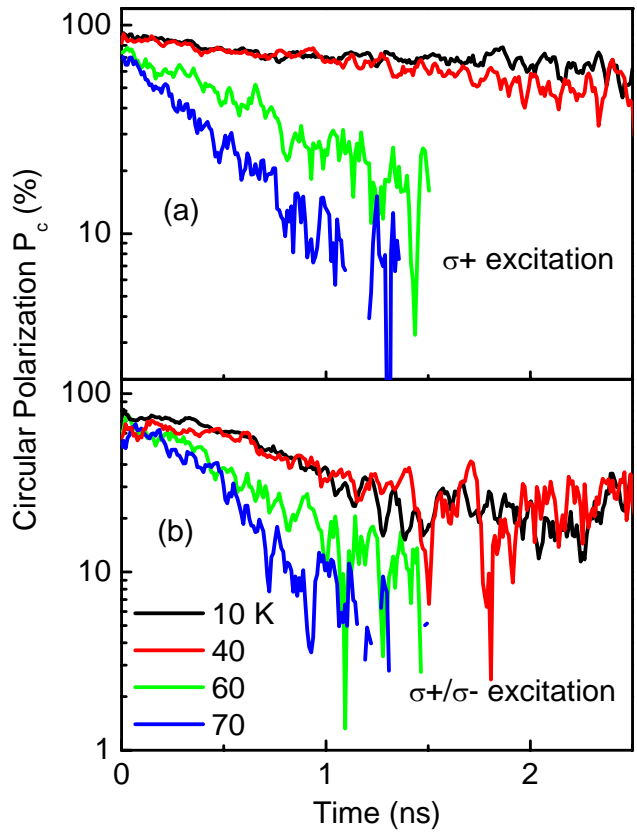

Figure 2

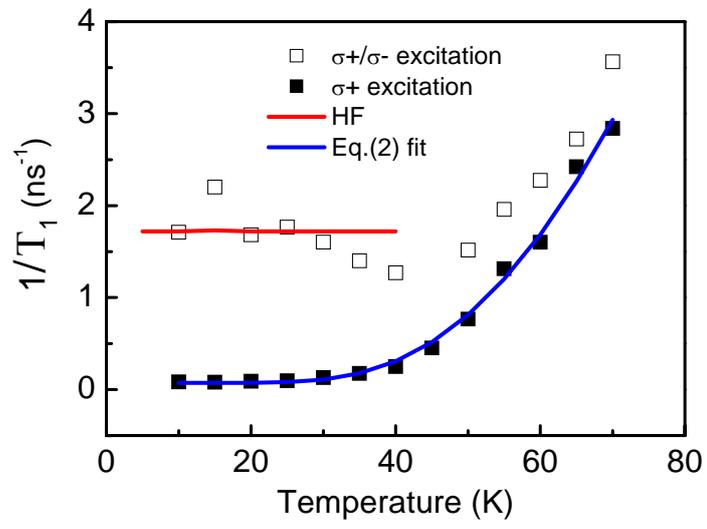

Figure 3

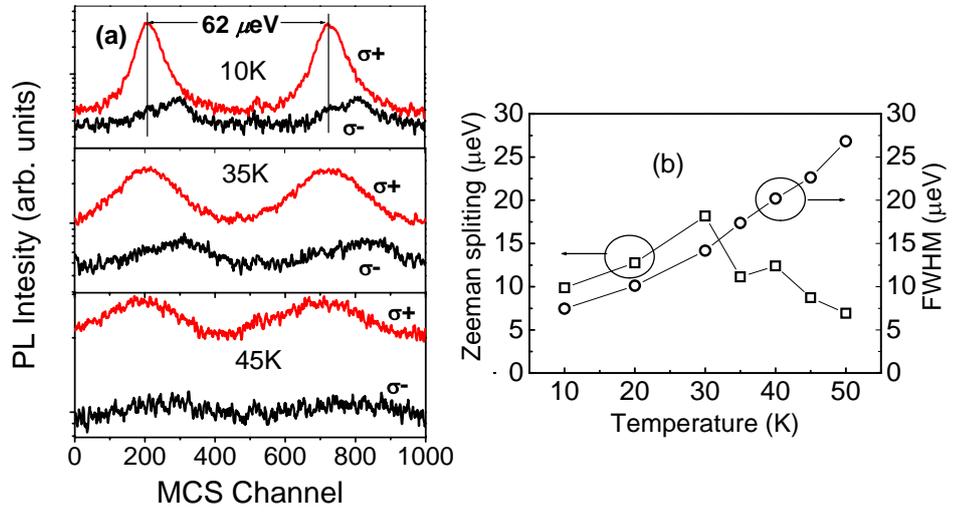

Figure 4